\documentclass[12pt,aps,eqsecnum,nofootinbib,floatfix]{revtex4}

\usepackage{graphics}
\usepackage{graphicx}
\usepackage{epstopdf}
\usepackage{subfigure}
\usepackage{amssymb,amsmath}
\usepackage{CJK}
\usepackage{indentfirst}
\usepackage{amsmath}

\def\be{\begin{equation}}
\def\ee{\end{equation}}
\def\bea{\begin{eqnarray}}
\def\eea{\end{eqnarray}}
\def\pt{\partial}
\def\nn{\nonumber}

\begin{document}

\title{Effects of global monopole on critical behaviors and microstructure of charged AdS black holes}

\author{Gao-Ming Deng$^{1}$\footnote{e-mail:~gmd2014cwnu@126.com}, Jinbo Fan$^{2}$, Xinfei Li$^{3}$}

\medskip

\affiliation{\footnotesize $^1$ School of Physics and Space Science, China West Normal University, Nanchong 637002, China\\
                           $^2$ National Demonstration Center for Experimental Physics and Electronics Education, Henan University, Kaifeng 475004, China\\
                           $^3$ School of Science, Guangxi University of Science and Technology, Guangxi 545026, China}

\begin{abstract}

As an intriguing topological defect, global monopole's influence on behaviors of black holes has always been anticipated but still remains less clear. Analyzing the thermodynamics of charged AdS black hole incorporating a global monopole manifests that the black hole undergoes a first-order phase transition at critical point, and of special interest, the critical behaviors qualitatively resemble a Van der Waals liquid-gas system. This paper concentrates on further investigating the charged AdS black hole with a global monopole, aiming at clarifying the significant effects of the global monopole on criticality and microstructure of charged AdS black holes. An interesting dependence on the internal global monopole can be witnessed intuitively by employing contrastive illustrations.
\bigskip

\textit{Keywords}: Global monopole;~AdS black holes;~critical behavior;~microstructure.

PACS numbers:{~04.70.-s,~04.70.Dy,~11.27.+d,~64.60.-i}

\end{abstract}

\maketitle

\tableofcontents
\bigskip

\section{Introduction} \label{section1}

Research on topological defects such as cosmic strings, domain walls, monopoles and textures has captured growing attention for several decades and still remains one of the most active fields in modern physics. It is widely believed that their fascinating properties which differ from those of more familiar systems can give rise to a rich variety of quite unusual physical phenomena \cite{Vilenkin:1984ib,Vilenkin:1994cs}. As an intriguing one, global monopoles \cite{Barriola:1989hx} arise from the phase transition of a system composed of a self-coupling triplet scalar field whose original global $O(3)$ gauge symmetry is spontaneously broken into $U(1)$. They usually carry a great many important messages about the Universe evolution in the early stage. An in-depth investigation of them has a great significance.

Promoted by the development of astronomical observation techniques, study of black holes in various spacetimes has received overwhelming concern \cite{Deng:2014bta,Deng:2017abh,Addazi:2016prb,Chen:2019nsr,Wei:2019ctz,Wei:2019uqg,Jusufi:2020odz,Li:2020vpo,Zeng:2020dco}. Especially, when a black hole swallows global monopole, properties of the black hole system become more interesting and deserve a deeper investigation. In 1989, Barriola and Vilenkin \cite{Barriola:1989hx} first discussed the gravitational field of a global monopole and creatively constructed a solution of Schwarzschild black hole with global monopole. Recently, black hole solution incorporating a global monopole has also been successfully generalized to the case of a charged black hole in asymptotically Anti-de Sitter (AdS) spacetime \cite{Deng:2018wrd}. Nowadays, the physical properties of black holes incorporating global monopoles have been detected extensively \cite{Yu:1994fy,Dadhich:1997mh,Gao:2002hf,Chen:2005vq,Ahmed:2016ucs,Jusufi:2017lsl,Rizwan:2018mpy,Das:2019khz,Secuk:2019njc}, and a great many valuable findings were harvested. However, how global monopoles influence the spacetime still remains less clear. Motivated by this fact, this present work will further explore the global monopole in the context of the charged AdS black hole \cite{Deng:2018wrd}, aiming at elaborating its significant effects on the charged AdS black holes, especially the critical behaviors and microstructure.

This paper proceeds as follows. In Sec.\ref{section2}, we briefly review the solution of charged AdS black hole \cite{Deng:2018wrd} with a global monopole and its thermodynamics. In Sec.\ref{section3}, the local stability and phase transitions are analyzed. We concentrate on exploring the heat capacity and exhibiting phase structures of the black hole. Sec.\ref{section4} is mainly devoted to investigating the characteristic critical behaviors. The significant impacts of global monopole on the criticality of the charged AdS black hole can be observed clearly. In Sec.\ref{section5}, we adopt thermodynamic geometry to further detect the phase transition and underlying properties of microstructure. This paper ends up with some discussions and conclusions in Sec.\ref{section6}.

\section{Black Hole Solution and Thermodynamics} \label{section2}

To facilitate further exploring the effects of global monopole on critical behaviors and microstructure of black holes, we first review the solution of charged AdS black hole with a global monopole and work out its thermodynamical quantities.

The gravitational field of a global monopole arising from a global $O(3)$ symmetry breaking was first discussed by Barriola and Vilenkin \cite{Barriola:1989hx}. The simplest model for generating such global monopoles can be described by the Lagrangian density:
\be
\mathcal{L}_{g m}=\frac{1}{2} \pt_{\mu} \phi^a \pt^{\mu} \phi^{\ast a} - \frac{\gamma}{4} {\left( \phi^a \phi^{\ast a} - \eta_0^2\right)}^2, \label{eq01}
\ee
in which $\phi^a$$=$$\eta_0 h(r) x^a/r$ is a triplet of scalar field, $\gamma$ a coupling constant, $\eta_0$ the energy scale of gauge symmetry breaking. For static spherically symmetric spacetime, considering the following ansatz
\be
d\tilde{s}^2=-\tilde{f}(\tilde{r}) d \tilde{t}^2+{\tilde{f}(\tilde{r})}^{-1} d \tilde{r}^2+\tilde{r}^2 d \Omega^2\,, \label{eq02}
\ee
Barriola and Vilenkin successfully constructed a Schwarzschild black hole solution with a global monopole, in which the metric function reads
\be
\tilde{f}(\tilde{r})=1 - 8 \pi \eta_0^2-\frac{2\tilde{m}}{\tilde{r}}\,. \label{eq03}
\ee
Recently, the solution has been generalized to the case of a charged black hole in asymptotically AdS spacetime \cite{Deng:2018wrd}. The line element of charged AdS black hole with a global monopole and gauged potential can be explicitly expressed as
\be
ds^2=-f(r)dt^2+\frac{1}{f(r)} dr^2 +\left(1-\eta ^2\right) r^2 \left(d\theta^2 + sin^2\theta d\phi^2\right)\,, \label{eq04}
\ee
\be
A=\frac{q}{r}dt\,,\quad
f(r)=1-\frac{2 m}{r}+\frac{q^2}{r^2}-\frac{\Lambda r^2}{3}\,, \label{eq05}
\ee
where $m$ and $q$ are, respectively, the mass parameter and electric charge parameter, the cosmology constant $\Lambda$ always takes negative value in asymptotically AdS spacetime. It is worth pointing out that this solution would reduce to standard four-demensional Reissner-Nordstr\"{o}m (RN) AdS black hole \cite{Ma:2016aat} if specifying the symmetry breaking parameter $\eta=0$. What's more, there a solid angle deficit emerges in the spacetime in comparison with the case of RN AdS black hole.

With the solution in hand, we proceed to work out thermodynamical quantities of the charged AdS black hole incorporating a global monopole. First of all, the outer horizon of the black hole is located at $r=r_h$ which is the largest root of equation $f(r)=0$. The electric charge and corresponding potential measured at infinity can be determined as \cite{Deng:2018wrd}
\bea
Q &=& \frac{1}{4 \pi} \oint F^{\mu \nu} d^2 \Sigma_{\mu \nu}=\left(1-\eta ^2\right)q\,, \\ \label{eq06}
\Phi &=& \xi^{\mu} A_{\mu} \big |_{r=r_h}=\frac{q}{r_h}\,, \label{eq07}
\eea
where $\xi^{\mu}$ is the timelike killing vector $(\partial _t)^{\mu}$. The ADM mass $M$ can be derived via the Komar integral
\be
M=\frac{1}{8\pi}\oint \xi_{(t)}^{\mu;\nu }d^2\Sigma_{\mu \nu }=(1-\eta^2)m\,. \label{eq08}
\ee
Resorting to the definition of the surface gravity, the Hawking temperature is given by
\be
T=\frac{f'(r_h)}{4 \pi}=\frac{\left(1-\eta^2\right)^2 r_h^2 \left(1-\Lambda r_h^2\right)-Q^2}{4\pi  \left(1-\eta^2\right)^2 r_h^3}\,. \label{eq09}
\ee
Moreover, we can also obtain the entropy of the black hole as follows
\be
S=\frac{1}{4} \int_{r=r_h}\sqrt{g_{\theta \theta}g_{\phi \phi}}~d\theta d\phi=\pi (1-\eta ^2) r_h^2\,. \label{eq10}
\ee
Obviously, when a charged AdS black hole swallows a global monopole, the thermodynamical quantities prove to be of great relevance to the symmetry breaking parameter $\eta$, they exhibit an interesting dependence on the internal global monopole.

\section{The Local Stability and Phase Transition} \label{section3}

\begin{figure*}[htbp]
  \begin{center}
    \subfigure[~$T$ versus $r_h$ for varying charge.]{\label{TRhdiag}\includegraphics[width=3.0in,height=2.3in]{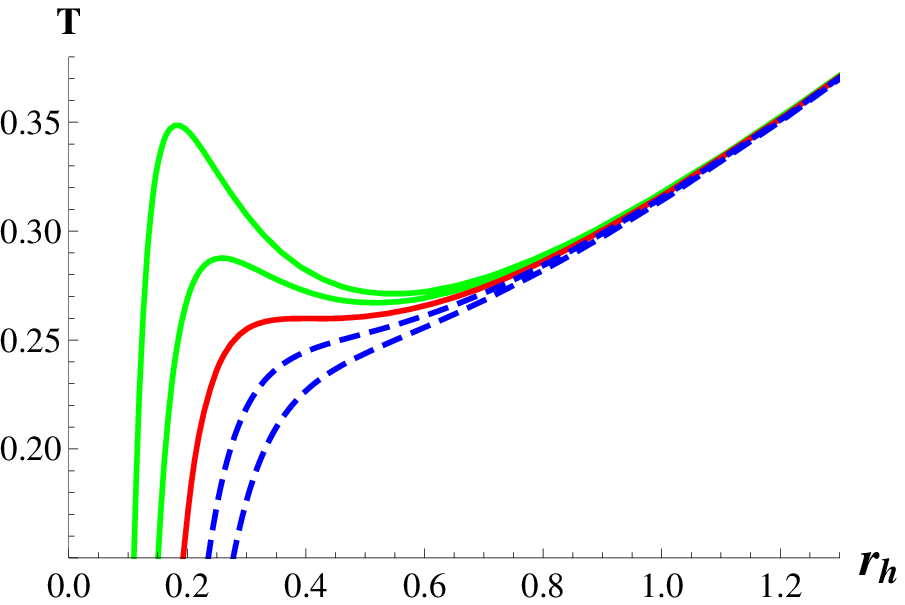}} \quad
    \subfigure[~$T$ versus $S$ for varying charge.]{\label{TSdiag}\includegraphics[width=3.0in,height=2.3in]{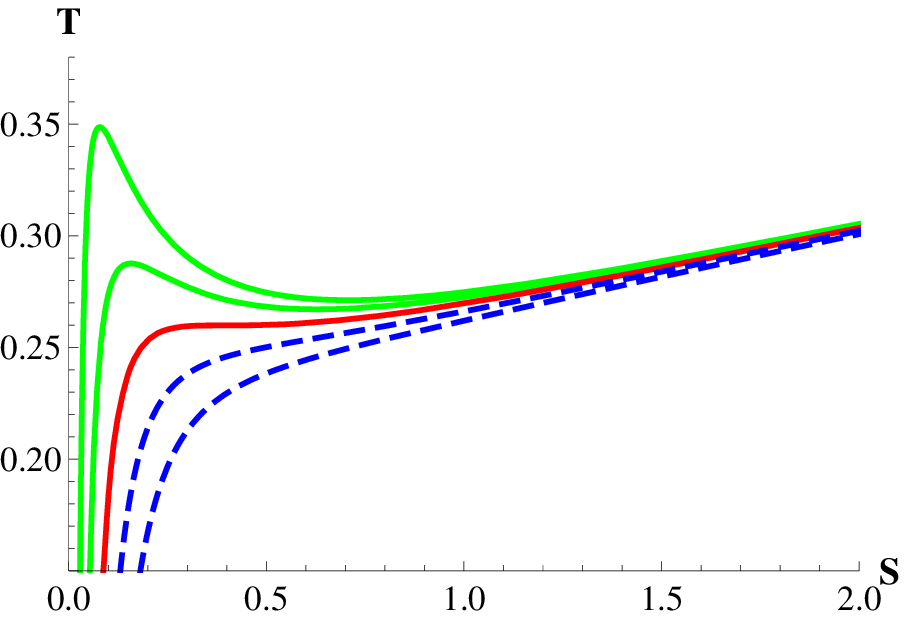}}
    \caption{(Color online) Behaviors of Hawking temperature. In each picture, the charge of isocharges decreases from bottom (blue dashed) to top (green solid). Here we set $\eta=0.5,\Lambda=-3$.}
  \label{TRhSdiag}
  \end{center}
\end{figure*}

Based on the thermodynamical quantities, we move to further analyze the attractive behaviors of the charged AdS black hole incorporating a global monopole.
To begin with, behaviors of the Hawking temperature defined in Eq. (\ref{eq09}) are vividly depicted in Fig.\ref{TRhSdiag}.
%%%%%%%%%%%%%%%%%%%%%%%%%%%%%%%%%%%%%%%%%%%%%%%%%%%%%%%%%%%%%%%
%%%%%%%%%%%%%%%%%%%%%%%%%%%%%%%%%%%%%%%%%%%%%%%%%%%%%%%%%%%%%%%
It should be noted that, within a certain range of temperature, top two (green solid) isocharges in each picture can be divided into three branches by isotherms, which correspond to three black hole solutions. The small and the large radius branches are thermodynamically stable, while the intermediate is unstable since the heat capacity $C$$=$$T\frac{\pt S}{\pt T}$ is plagued by negative values. For the middle (red solid) isocharge, an inflection point is exhibited. For the bottom (blue dashed) isocharges, the Hawking temperature increases monotonically, there may exist only one black hole solution. Consequently, the behaviors summarized in Fig.\ref{TRhSdiag} might have a subtle hint for phase transition.

\begin{figure*}[htbp]
  \begin{center}
    \subfigure[~$C_Q$ for the charged AdS black hole with a global monopole ($\eta$=$0.5$).]{\label{CQRhQ3D05diag}\includegraphics[width=3.0in,height=2.3in]{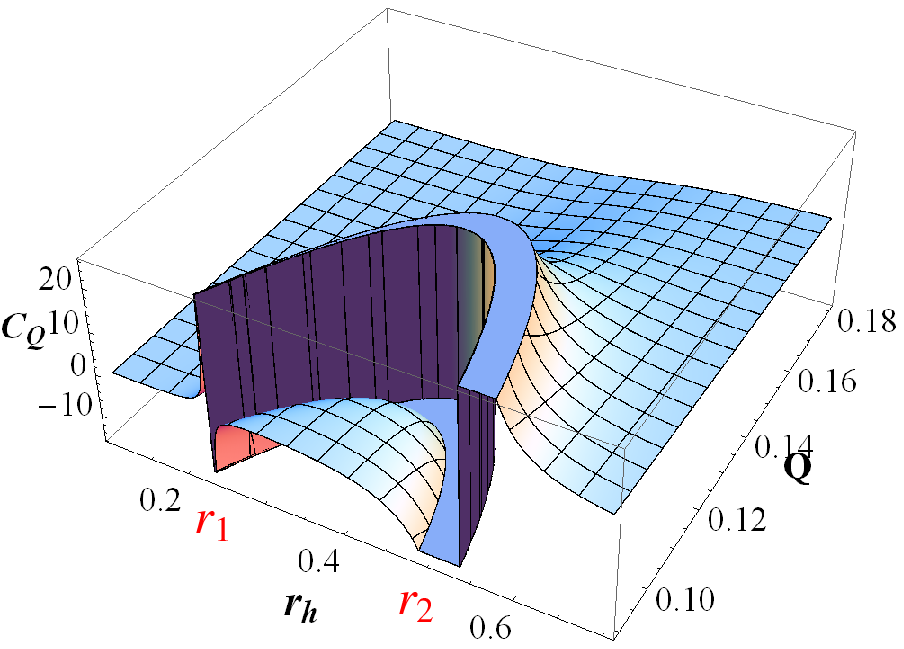}} \quad
    \subfigure[~$C_Q$ for the standard Reissner-Nordstr\"{o}m AdS black hole ($\eta$=$0$).]{\label{CQRhQ3D00diag}\includegraphics[width=3.0in,height=2.3in]{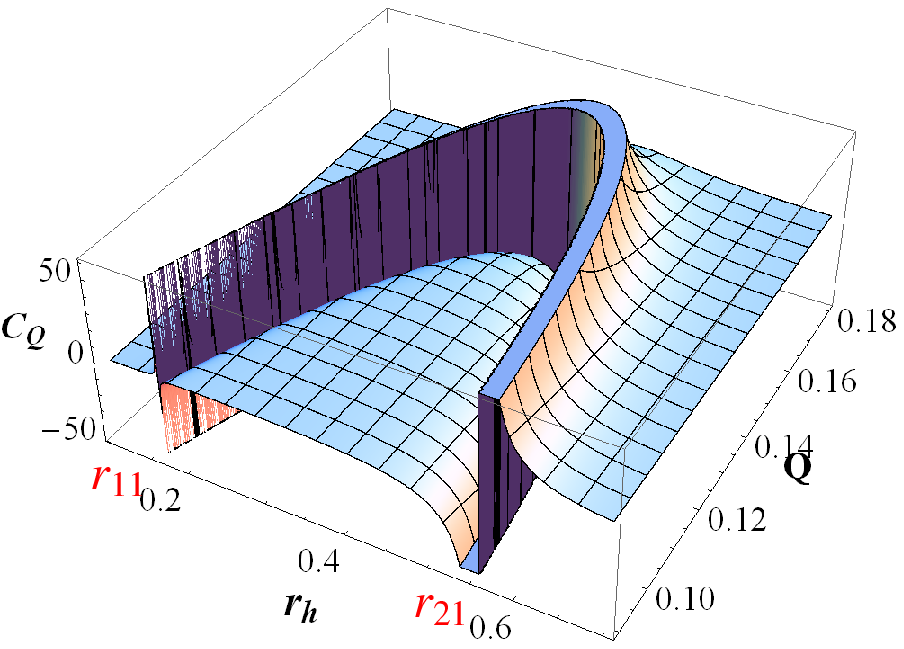}}
    \caption{(Color online) Heat capacity $C_Q$ varying with $r_h$ and $Q$ for the charged AdS black hole with a global monopole and the standard Reissner-Nordstr\"{o}m AdS black hole. Here $\Lambda$=$-3$ is fixed.}
  \label{CQRhQ3Ddiag}
  \end{center}
\end{figure*}

\begin{figure*}[htbp]
  \begin{center}
    \subfigure[~$C_Q$ versus $r_h$, $Q=0.09<Q_c$, $\eta=0.5$.]{\label{CQRh2D0501diag}\includegraphics[width=2.6in,height=2.2in]{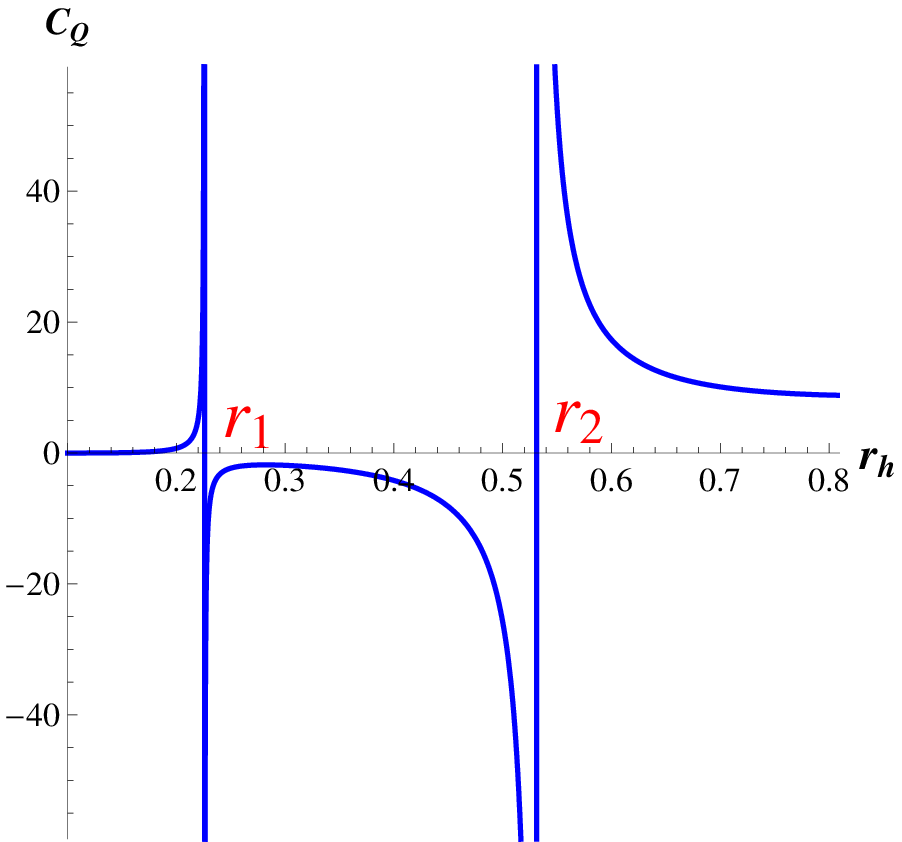}} \qquad
    \subfigure[~$C_Q$ versus $r_h$, $Q=0.09<Q_{c1}$, $\eta=0$.]{\label{CQRh2D0001diag}\includegraphics[width=2.6in,height=2.2in]{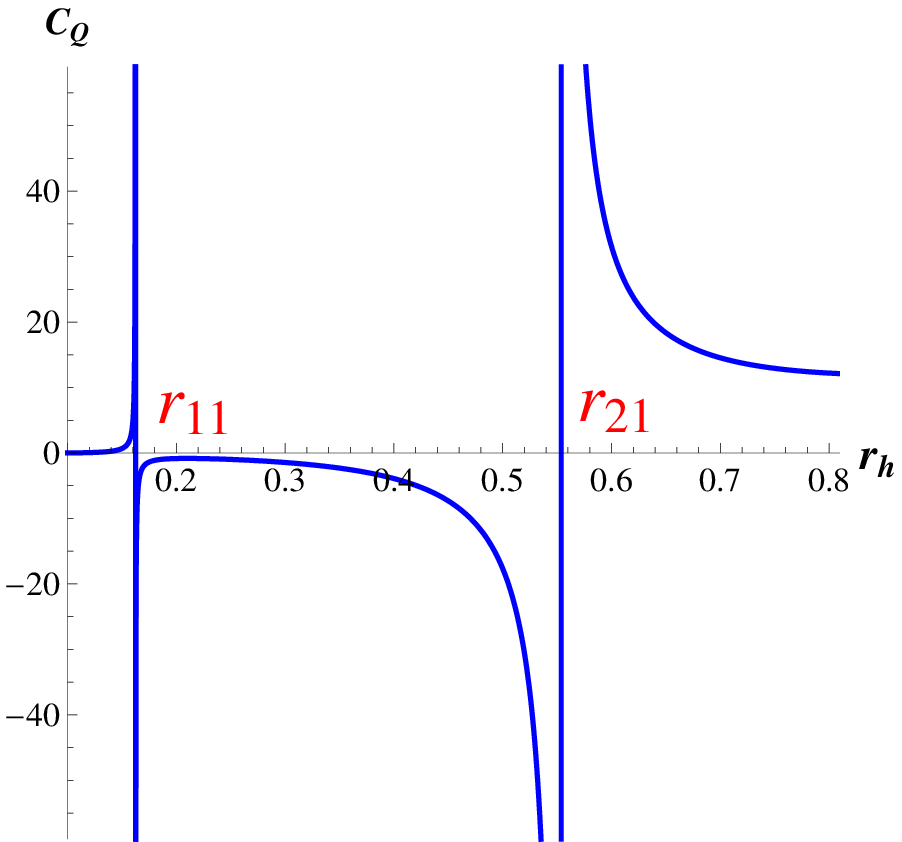}} \\
    \subfigure[~$C_Q$ versus $r_h$, $Q=Q_c$, $\eta=0.5$.]{\label{CQRh2D0502diag}\includegraphics[width=2.6in,height=2.2in]{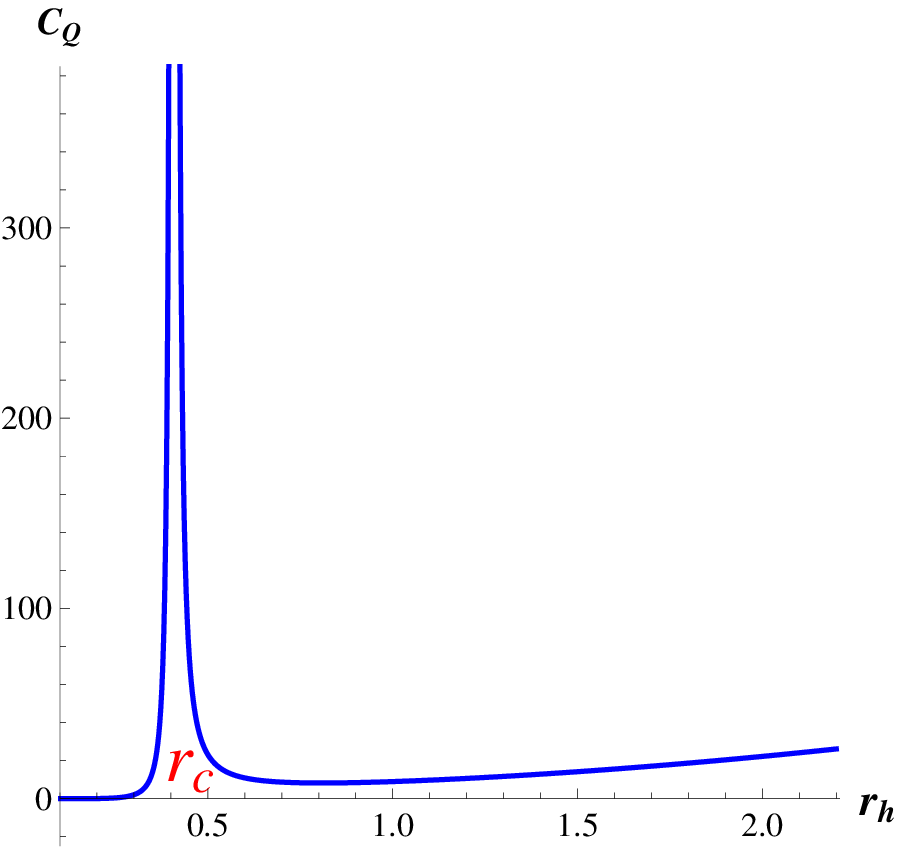}} \qquad
    \subfigure[~$C_Q$ versus $r_h$, $Q=Q_{c1}$, $\eta=0$.]{\label{CQRh2D0002diag}\includegraphics[width=2.6in,height=2.2in]{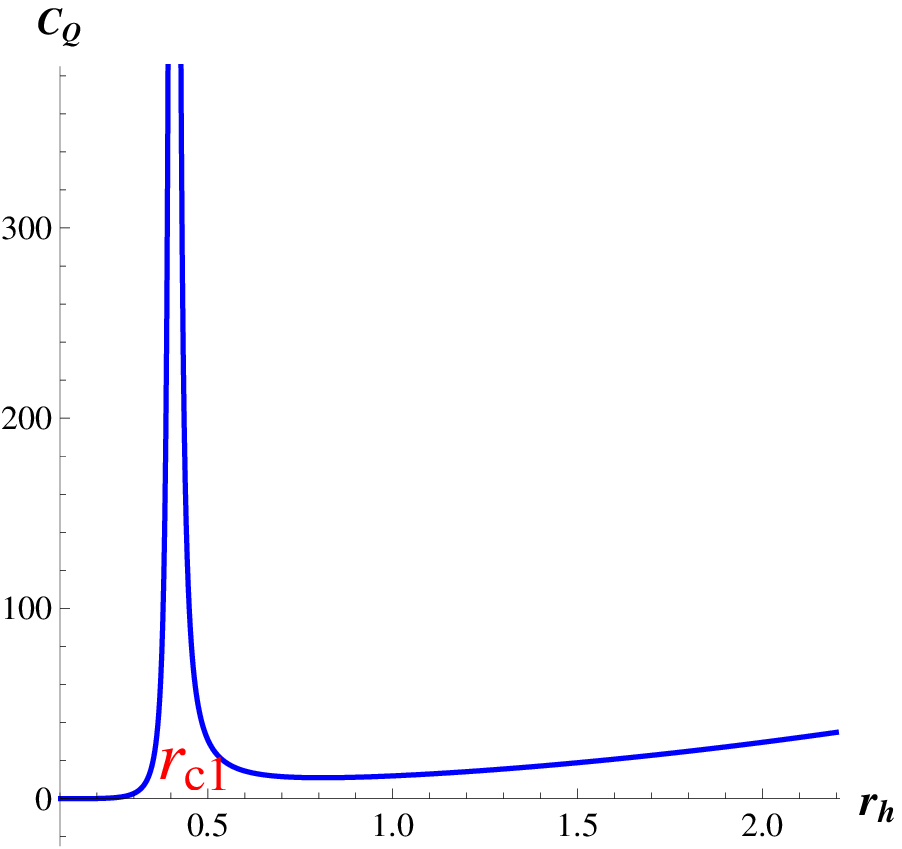}}\\
    \subfigure[~$C_Q$ versus $r_h$, $Q=0.25>Q_c$, $\eta=0.5$.]{\label{CQRh2D0503diag}\includegraphics[width=2.6in,height=2.2in]{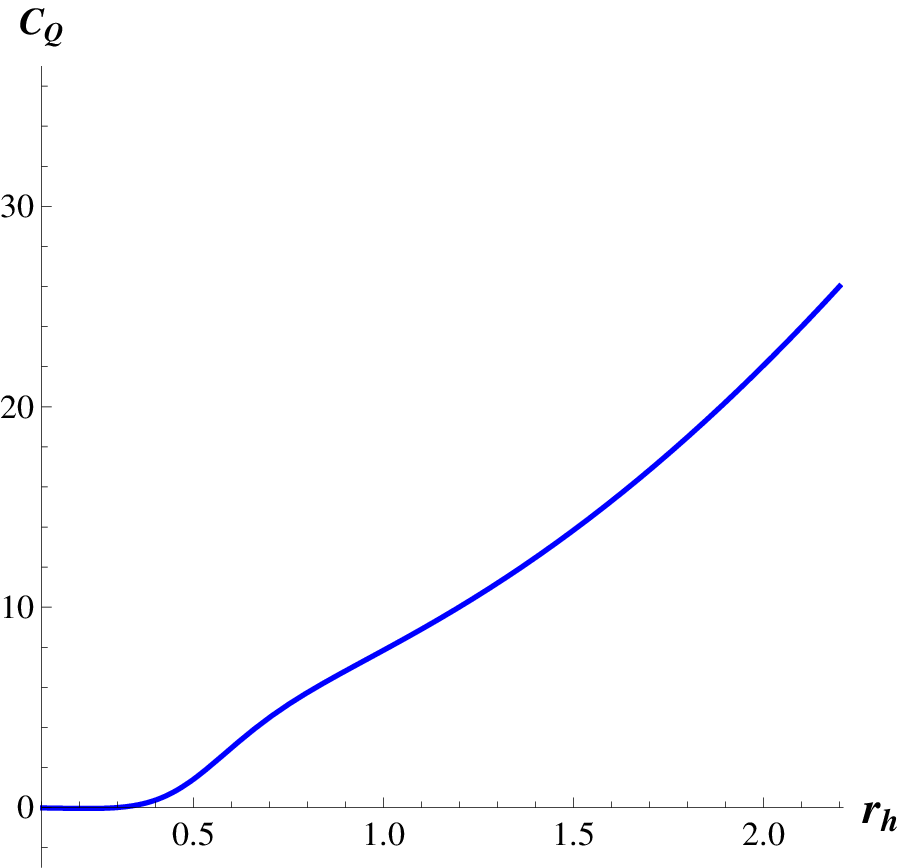}} \qquad
    \subfigure[~$C_Q$ versus $r_h$, $Q=0.25>Q_{c1}$, $\eta=0$.]{\label{CQRh2D0003diag}\includegraphics[width=2.6in,height=2.2in]{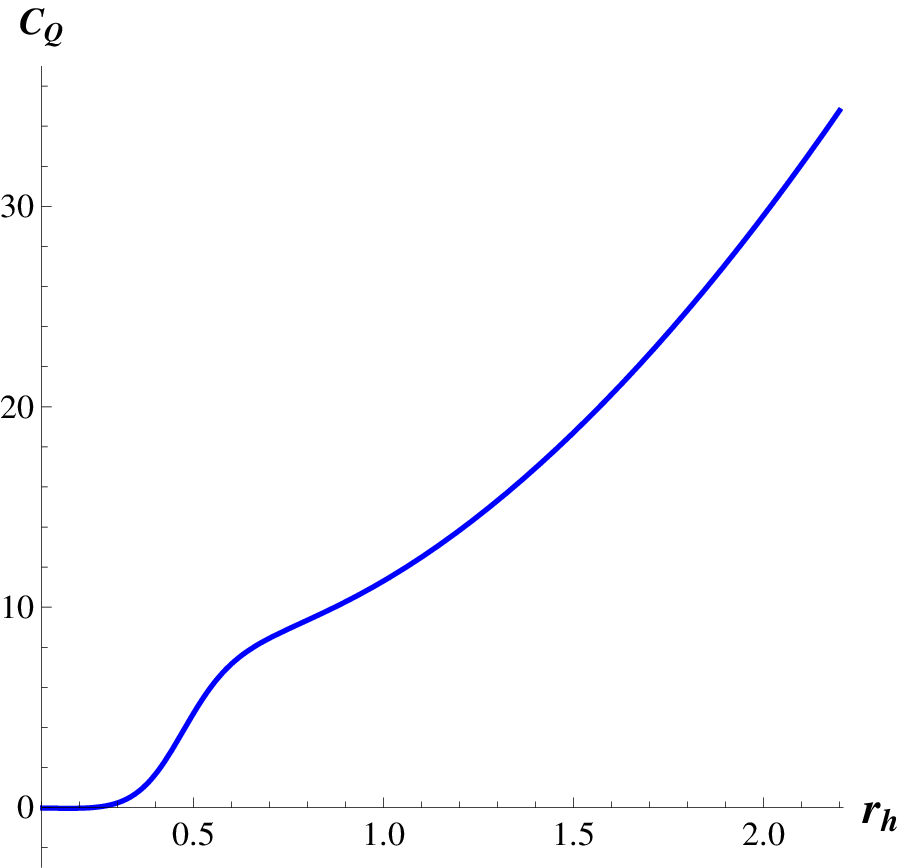}}
    \caption{(Color online) The divergent behaviors of the heat capacity $C_Q$. The left three pictures (a), (c) and (e) correspond to the charged AdS black hole incorporating a global monopole, the right three pictures (b), (d) and (f) correspond to the Reissner-Nordstr\"{o}m AdS black hole ($\eta$=$0$). Here we set $\Lambda$=$-3$.}
  \label{CQRh2Ddiag}
  \end{center}
\end{figure*}

In fact, a signal of phase transition also can be decoded from the heat capacity. In general, a positive heat capacity ensures a stably existing black hole, while a negative one signals that the black hole will disappear when encountering a slight perturbation. With Eqs. (\ref{eq09}) and (\ref{eq10}) in hand, the heat capacity $C_Q$ at constant charge can be derived as
\be
C_Q = T \left(\frac{\pt S}{\pt T} \right)_{Q} = \frac{2 \pi \left(1-\eta^2\right) r_h^2 \left[\left(1-\eta^2\right)^2 r_h^2 \left(1-\Lambda r_h^2\right)-Q^2\right]}{-\left(1-\eta^2\right)^2 r_h^2 \left(1+\Lambda r_h^2\right)+3 Q^2}\,. \label{eq11}
\ee
Undoubtedly, the heat capacity $C_Q$ will diverge if the denominator vanishes, that is
\be
-\left(1-\eta^2\right)^2 r_h^2 \left(1+\Lambda r_h^2\right)+3 Q^2 = 0\,. \label{eq12}
\ee
Solutions for $r_h$ degenerate into single one when
\be
\left(1-\eta^2\right)^2+12 \Lambda  Q^2=0\,. \label{eq13}
\ee
The corresponding electric charge and horizon radius turn out to be
\be
Q_c = \frac{1-\eta^2}{2 \sqrt{-3 \Lambda}}\,, \quad r_{c} = \frac{1}{\sqrt{-2 \Lambda}}\,. \label{eq14}
\ee
One may appreciate behaviors of the heat capacity $C_Q$ changing with $r_h$ and $Q$ in Fig.\ref{CQRhQ3Ddiag}.
%%%%%%%%%%%%%%%%%%%%%%%%%%%%%%%%%%%%%%%%%%%%%%%%%%%%%%%%%%%%%%%
%%%%%%%%%%%%%%%%%%%%%%%%%%%%%%%%%%%%%%%%%%%%%%%%%%%%%%%%%%%%%%%
Apparently, the heat capacity $C_Q$ suffers from a striking divergence at the points $r_1$ and $r_2$. Closer observations can be acquired in Fig.\ref{CQRh2Ddiag}.
%%%%%%%%%%%%%%%%%%%%%%%%%%%%%%%%%%%%%%%%%%%%%%%%%%%%%%%%%%%%%%%
%%%%%%%%%%%%%%%%%%%%%%%%%%%%%%%%%%%%%%%%%%%%%%%%%%%%%%%%%%%%%%%
In detail, for a small value of the electric charge $Q<Q_c$, Fig.\ref{CQRh2D0501diag} displays three regions partitioned by two divergent points $r_1$ and $r_2$. $C_Q$ remains positive for regions $r_h<r_1$ and $r_h>r_2$, while it is negative for the region $r_1<r_h<r_2$. It means that both the small radius region $r_h<r_1$ and the large radius region $r_h>r_2$ are thermodynamically stable while the intermediate radius region $r_1<r_h<r_2$ is unstable. As a result, the intermediate black hole (IBH) can not exist stably, there will indeed be a phase transition occurs between small black hole (SBH) and large black hole (LBH). With the charge increasing to $Q=Q_c$, two divergent points approach to each other and ultimately merge into one at the point $r_h=r_c$ (see Fig.\ref{CQRh2D0502diag}). When the charge $Q>Q_c$ (see Fig.\ref{CQRh2D0503diag}), the fact that $C_Q$ always remains positive reports that the black hole is locally stable and no phase transition will be triggered. We plot the heat capacity $C_Q$ of the standard Reissner-Nordstr\"{o}m AdS black hole in Fig.\ref{CQRhQ3D00diag} and Figs.\ref{CQRh2D0001diag}, \ref{CQRh2D0002diag} and \ref{CQRh2D0003diag} for comparison.
% global monopole's impact on behaviors of the black hole coincides with

\section{Van der Waals-like Critical Behaviors} \label{section4}

Analyzing the stability of the charged AdS black hole with a global monopole reveals that the black hole really suffers from a small-large black hole phase transition. In what follows, we step forward to give insight into the critical behaviors and focus on reporting impacts of global monopole on the black hole.

We start with expressing $Q$ as a function of $\Phi$ and $T$, namely
\be
Q = \frac{\left(1-\eta^2\right)\Phi}{-\Lambda }\left[2 \pi  T-\sqrt{4 \pi^2 T^2+\Lambda  \left(1-\Phi^2\right)}\right]\,, \label{eq15}
\ee
where $\eta$ is the symmetry breaking parameter characterizing the global monopole, $\Lambda$ the cosmology constant which takes negative values in asymptotically AdS spacetime. Combining the equation of state (\ref{eq15}) and the derivatives
\be
\left(\frac{\pt Q}{\pt \Phi}\right)_T = 0 \, ,\quad \quad \left(\frac{\pt^2 Q}{\pt \Phi^2}\right)_T = 0\,,    \label{eq16}
\ee
the critical quantities can be written as
\be
T_c = \frac{\sqrt{-\Lambda}}{3 \pi} \, ,\quad Q_c = \frac{1-\eta^2}{2\sqrt{-3 \Lambda}}\,, \quad \Phi_c = \frac{1}{\sqrt{6}}\,. \label{eq17}
\ee
The result coincides with Eq.(\ref{eq14}). It is particularly interesting that, in the presence of a global monopole, parameter $\eta$ plays a decisive role in determining the critical point (\ref{eq17}).

\begin{figure*}[htbp]
  \begin{center}
    \mbox{
      \subfigure[~$\eta$=$0.5$.]{\label{QPhi0501diag}\includegraphics[width=2.6in,height=2.2in]{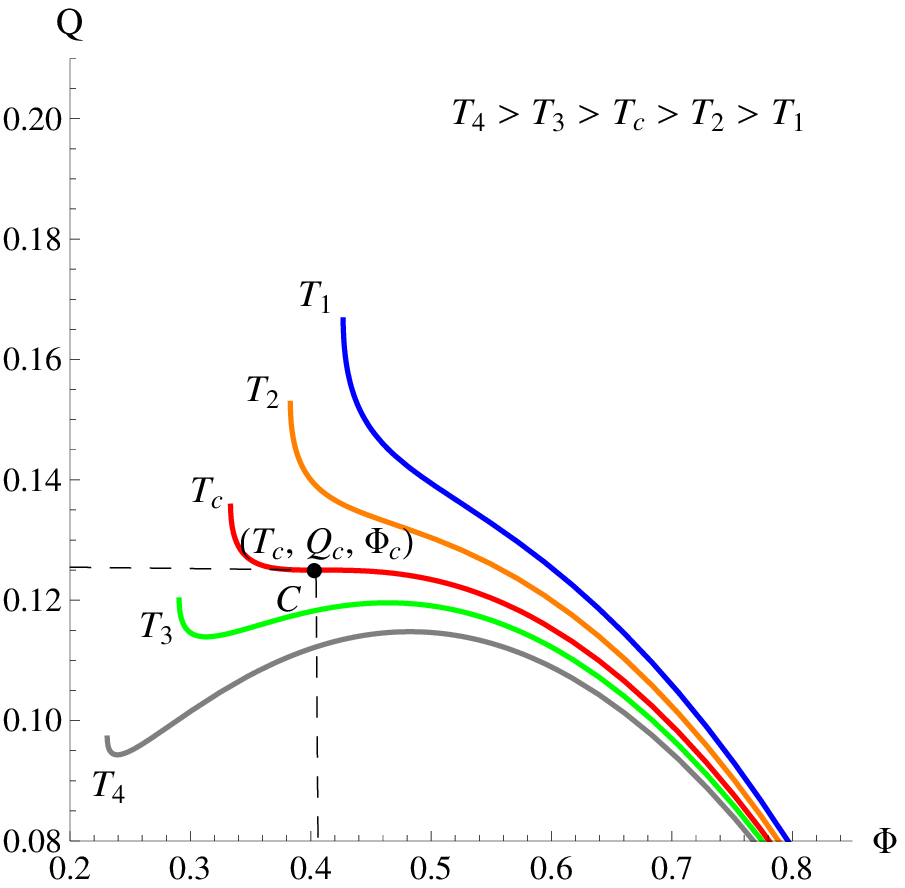}} \qquad
      \subfigure[~$\eta$=$0$.]{\label{QPhi0001diag}\includegraphics[width=2.6in,height=2.2in]{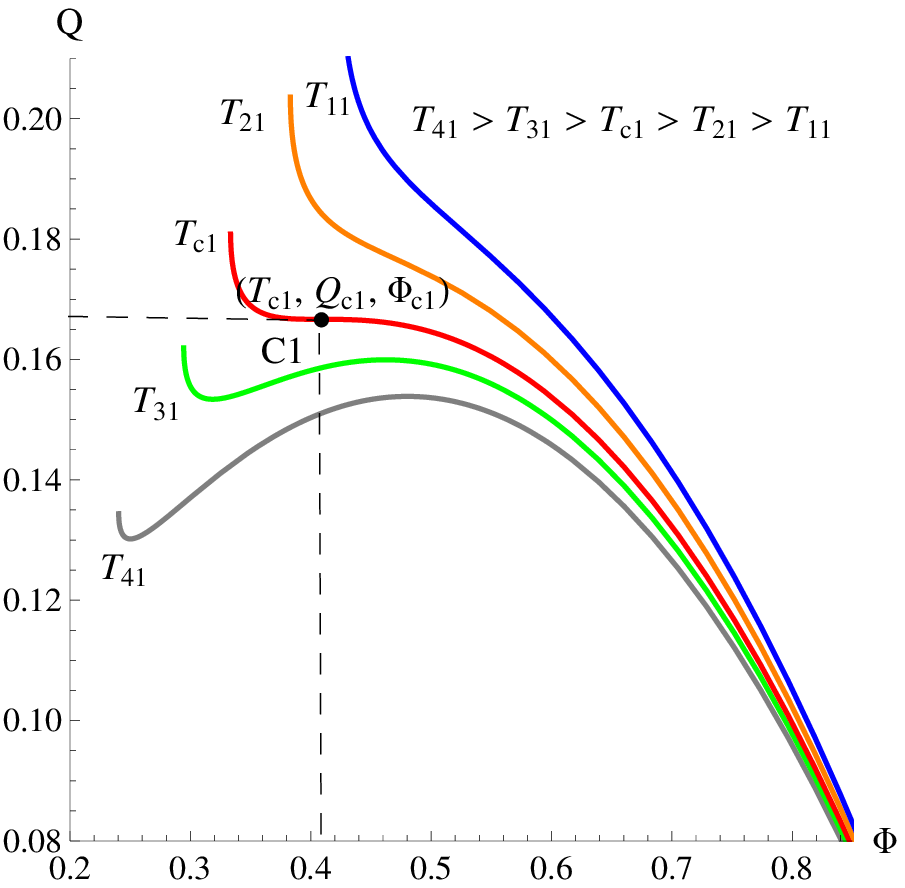}}
         }
    \mbox{
      \subfigure[~Closeup of the isotherm $T_3$ ($\eta=0.5$).]{\label{QPhicloseup0501diag}\includegraphics[width=2.6in,height=2.2in]{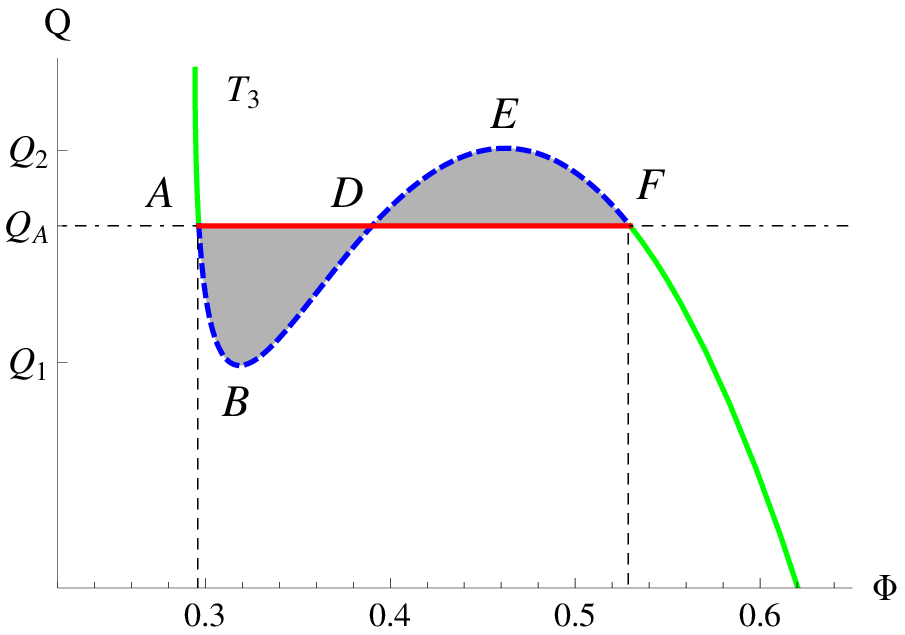}} \qquad
      \subfigure[~Closeup of the isotherm $T_{31}$ ($\eta=0$).]{\label{QPhicloseup0001diag}\includegraphics[width=2.6in,height=2.2in]{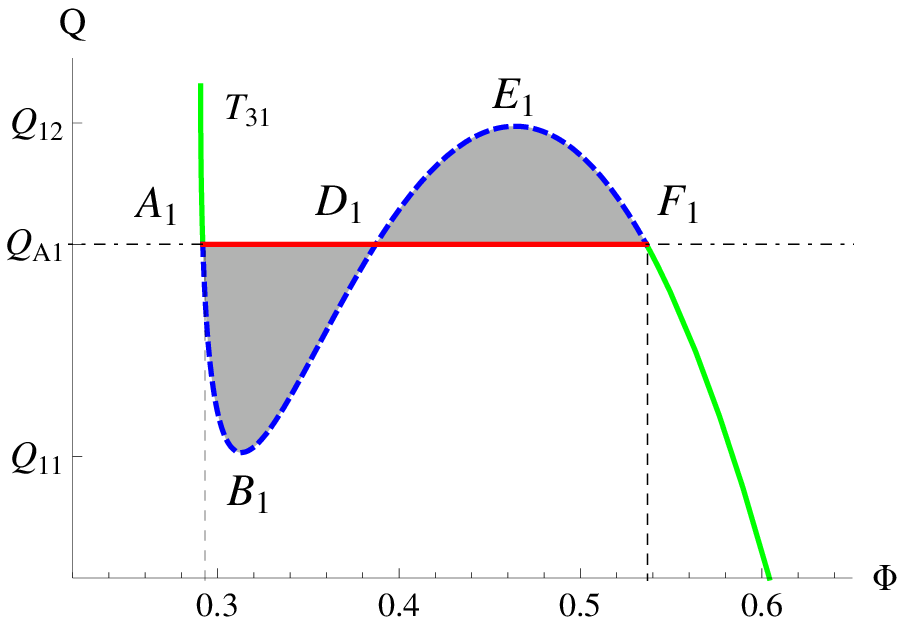}}
         }
    \caption{(Color online) $Q$-$\Phi$ diagram (isotherm) for different $\eta$. Critical entities are labelled with subscript $``c"$. Temperature of isotherms in each picture decreases from bottom to top. The critical isotherm is plotted by the red solid curve and tangent to isocharge line at the critical point. The limit case $\eta=0$ in the right two pictures (b) and (d) correspond to the well-known four-dimensional RN AdS black hole \cite{Ma:2016aat}. Here we set $\Lambda$=$-3$.}
  \label{QPhi2Ddiag}
  \end{center}
\end{figure*}

The qualitative behaviors of isotherms in $Q$-$\Phi$ plane can be observed clearly in Fig.\ref{QPhi2Ddiag}.
%%%%%%%%%%%%%%%%%%%%%%%%%%%%%%%%%%%%%%%%%%%%%%%%%%%%%%%%%%%%%%%
%%%%%%%%%%%%%%%%%%%%%%%%%%%%%%%%%%%%%%%%%%%%%%%%%%%%%%%%%%%%%%%
Seen from Figs.\ref{QPhi0501diag} and \ref{QPhicloseup0501diag}, the charged AdS black hole incorporating a global monopole exhibits eye-catching Van der Waals-like behaviors. When the temperature is higher than the critical value, the (green and gray solid) isotherms are characterized by the intriguing Van der Waals oscillation. With the help of Maxwell's equal area law, the unphysical oscillation curve segment shown in Fig.\ref{QPhicloseup0501diag} can be reconciled by replacing it with the isocharge line segment which the shadow area below and above are equal. Phase transition between a SBH and a LBH occurs along the refined physical isotherm. With temperature dropping to the critical value, the oscillating segment squeezes and eventually develops into the critical point. Meanwhile, the isotherm branch advances as the critical isotherm $T=T_c$. For the case $T<T_c$, there is no local minimum or maximum any longer, and $Q$ decreases monotonically with increasing $\Phi$.

Remarkably, Fig.\ref{QPhi0501diag} also displays how global monopole influences the critical behaviors of the charged AdS black hole. On the one hand, comparing Figs.\ref{QPhi0501diag} and \ref{QPhi0001diag}, it is evident that the appearance of a global monopole results in a lower critical point. Actually, the reason for that can be enlightened in Eq. (\ref{eq17}). In the presence of a global monopole, the critical quantities show a strong dependence on the symmetry breaking parameter $\eta$ for a given cosmology constant. Especially, the critical electric charge decreases with the parameter $\eta$ increasing. On the other hand, it is quite interesting to find that, the large $\eta$ promotes the black hole to reach the stable stage. Comparing the isotherms in Figs.\ref{QPhicloseup0501diag} and \ref{QPhicloseup0001diag} carefully, as the parameter $\eta$ increases, the unstable stage becomes shorter.  Therefore, we can infer that, the larger value the parameter $\eta$ takes, the easier the black hole system reaches the stable stage during phase transition.

This section mainly analyzed the critical behaviors of the charged AdS black hole with a global monopole. By contrast, we found that a global monopole has a notable influence on the behaviors of the black hole during phase transition. In the presence of a global monopole, the critical quantities become so closely dependent on the parameter $\eta$ that the critical point shifts downwards. Besides, more importantly, analysis above shows that
a global monopole plays a significant role in promoting the black hole to reach the stable stage when suffering from a phase transition.

\section{Thermodynamic Geometry and Microscopic Properties} \label{section5}

A systematic statistical description of black hole microstates has not been well-defined yet. Luckily, thermodynamic geometry is known as an elegant measure of global fluctuations in thermodynamic systems caused by interactions and available to give insight into the underlying statistical mechanical system \cite{Ruppeiner:1983zz,Ruppeiner:1995zz,Oshima:1999rs,Ruppeiner:2018pgn}. It has been advanced as a useful tool for studying microscopic properties of black hole system and been extensively adopted \cite{Mo:2013sxa,Mo:2015bta,Quevedo:2015fga,Soroushfar:2016nbu,L.:2018jss,Soroushfar:2019ihn,Ghosh:2019pwy,Yerra:2020oph,Kumara:2020ucr}. According to Ruppeiner's proposal, the thermodynamic geometry metric is defined as
\be
g_{i j}^R = -\frac{\pt^{2}S(x)}{\pt x^{i}\pt x^{j}}\,, \label{eq18}
\ee
where $x^{i}$ denotes extensive variables of a given thermodynamic system. Since Ruppeiner geometry is conformally related to Weinhold geometry \cite{Weinhold:1975mg,Mrugala:1983ct}, one usually expresses Ruppeiner metric resorting to Weinhold's definition for convenience, that is
\be
g_{i j}^R = \frac{1}{T}\frac{\pt^{2}U}{\pt x^{i}\pt x^{j}}\,, \label{eq19}
\ee
in which $T$ and $U$ are the temperature and internal energy of the black hole system respectively.

\begin{figure*}[htbp]
  \begin{center}
    \subfigure[~$\eta$=$0.5$.]{\label{RRhQ3D05diag}\includegraphics[width=2.8in,height=2.3in]{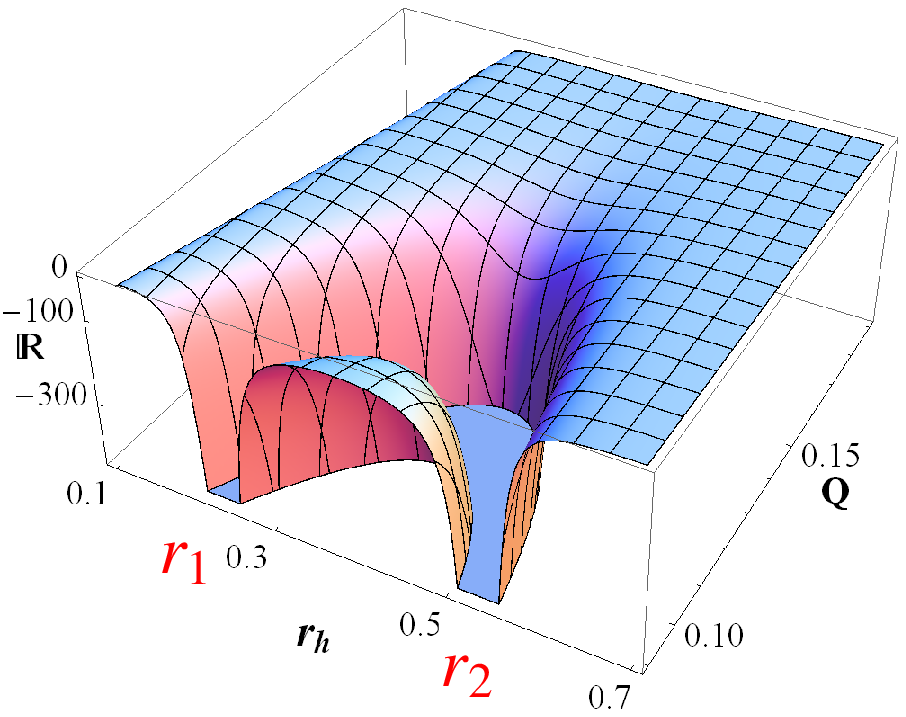}} \quad
    \subfigure[~$\eta$=$0$.]{\label{RRhQ3D00diag}\includegraphics[width=2.8in,height=2.3in]{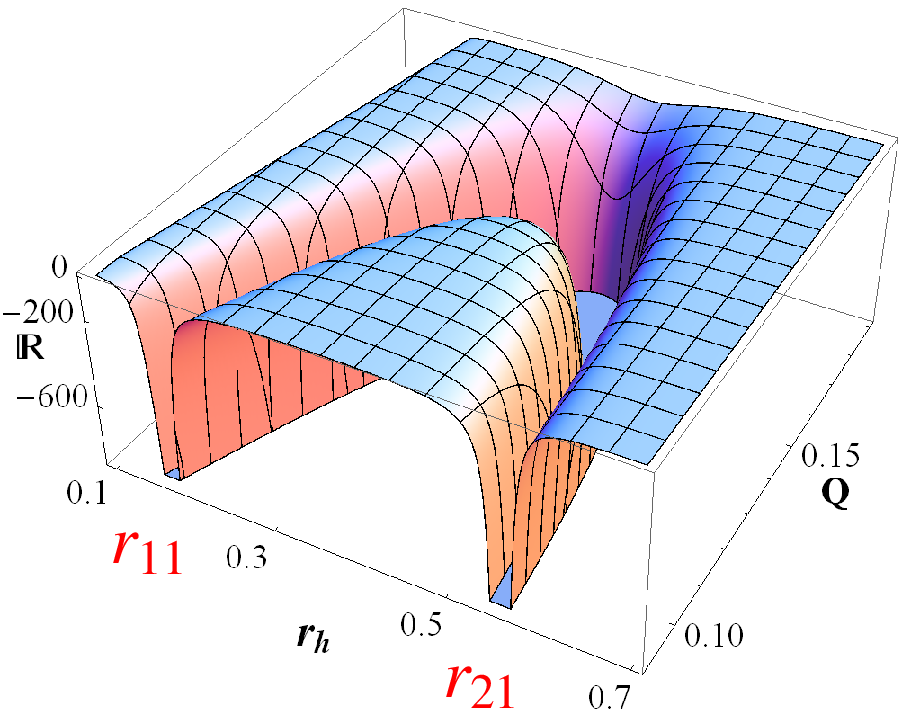}}
    \caption{(Color online) Ruppeiner curvature $\mathbb{R}$ varying with $r_h$ and $Q$. The left picture ($\eta=0.5$) corresponds to the charged AdS black hole with a global monopole, the right picture ($\eta=0$) corresponds to the Reissner-Nordstr\"{o}m AdS black hole. The case for extremal black holes is excluded. Here $\Lambda=-3$ is fixed.}
  \label{RRhQ3Ddiag}
  \end{center}
\end{figure*}

\begin{figure*}[htbp]
  \begin{center}
    \subfigure[~$\mathbb{R}$ versus $r_h$ for $\eta=0.5$.]{\label{RRh2D05diag}\includegraphics[width=2.8in,height=2.5in]{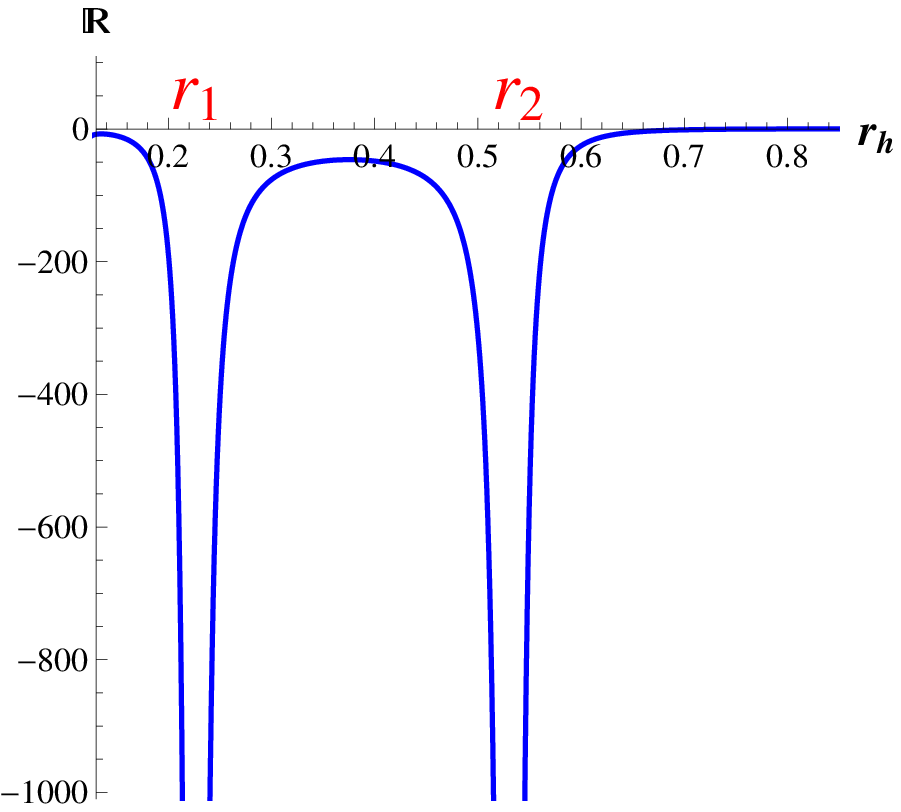}} \qquad
    \subfigure[~$\mathbb{R}$ versus $r_h$ for $\eta=0$.]{\label{RRh2D00diag}\includegraphics[width=2.8in,height=2.5in]{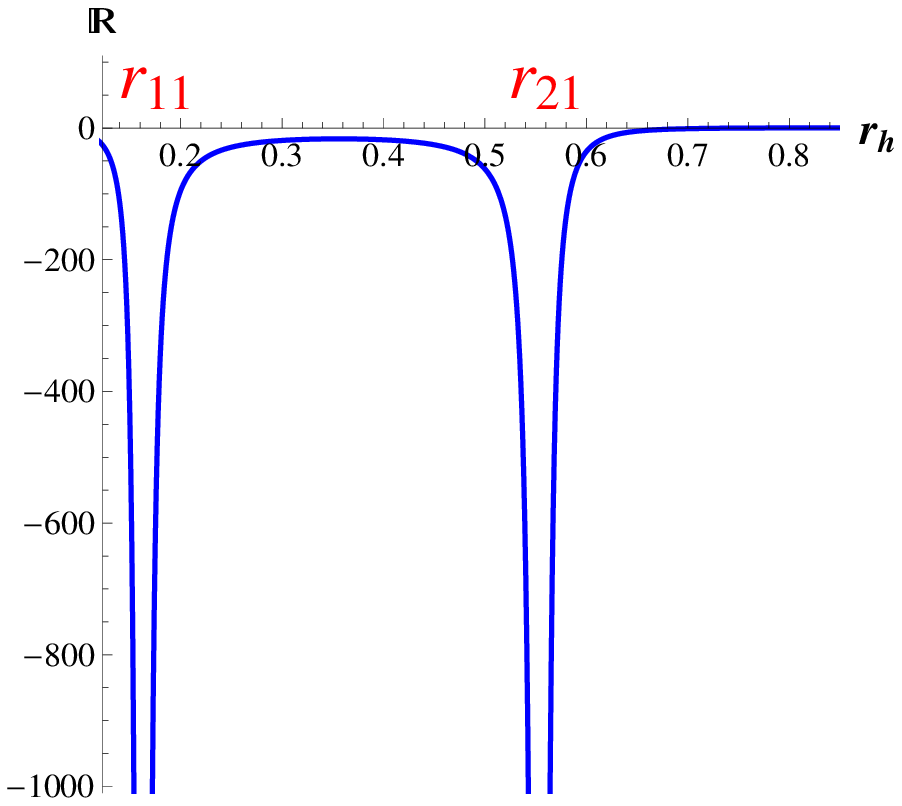}} \\
    \caption{(Color online) The divergent behaviors of Ruppeiner curvature $\mathbb{R}$. The left picture ($\eta=0.5$) corresponds to the charged AdS black hole with a global monopole, the right picture ($\eta=0$) corresponds to the Reissner-Nordstr\"{o}m AdS black hole. Here we set $\Lambda=-3$,\,$Q=0.09$.}
  \label{RRh2Ddiag}
  \end{center}
\end{figure*}

For the charged AdS black hole with a global monopole, the Ruppeiner metric can be given by
\bea
g_{SS}^R &=& \frac{\Lambda S+\pi \left(1-\eta^2\right) \left(1-\Phi^2\right)}{2 S \left[\Lambda S-\pi \left(1-\eta^2\right) \left(1-\Phi^2\right)\right]}\,, \label{eq20} \\
%g_{SS}^R &=& \frac{4 \pi  \left(\xi^2-\Phi ^2\right)+\xi R_0 S}{2 S \left(4 \pi  \left(-\xi^2+\Phi ^2\right)+\xi R_0 S\right)}\,, \label{eq20} \\
g_{S\Phi}^R &=& \frac{-2 \pi \left(1-\eta^2\right) \Phi}{\Lambda S-\pi \left(1-\eta^2\right) \left(1-\Phi^2\right)} = g_{\Phi S}^R\,, \label{eq21} \\
g_{\Phi \Phi}^R &=& \frac{4 \pi \left(1-\eta^2\right) S}{\Lambda S-\pi \left(1-\eta^2\right) \left(1-\Phi^2\right)}\,. \label{eq22}
\eea
Programming with Mathematica 12, we can compute the Ruppeiner scalar curvature as
\be
\mathbb{R}=\frac{\left(1-\eta^2\right) \Delta }{\pi  \left[\left(1-\eta^2\right)^2 r_h^2 \left(1-\Lambda r_h^2\right)-Q^2\right] \left[-\left(1-\eta^2\right)^2 r_h^2 \left(1+\Lambda r_h^2\right)+3 Q^2\right]^2}, \label{eq23}
\ee
where
\be
\Delta = \Lambda r_h^2 \left\{10 Q^4-\left(1-\eta^2\right)^2 r_h^2 \left(3+2 \Lambda r_h^2\right) \left[3 Q^2-\left(1-\eta^2\right)^2 r_h^2\right]\right\}-\left[Q^2-\left(1-\eta^2\right)^2 r_h^2\right]^2\,. \nn
\ee
It's easy to see that, considering the nonextremal condition (\ref{eq09}), scalar curvature (\ref{eq23}) shares the divergent points with heat capacity (\ref{eq11}). The divergent behaviors are portrayed in Figs.\ref{RRhQ3Ddiag} and \ref{RRh2Ddiag}.
%%%%%%%%%%%%%%%%%%%%%%%%%%%%%%%%%%%%%%%%%%%%%%%%%%%%%%%%%%%%%%%
%%%%%%%%%%%%%%%%%%%%%%%%%%%%%%%%%%%%%%%%%%%%%%%%%%%%%%%%%%%%%%%
%%%%%%%%%%%%%%%%%%%%%%%%%%%%%%%%%%%%%%%%%%%%%%%%%%%%%%%%%%%%%%%
%%%%%%%%%%%%%%%%%%%%%%%%%%%%%%%%%%%%%%%%%%%%%%%%%%%%%%%%%%%%%%%
Observing clearly from Fig.\ref{RRh2D05diag}, the Ruppeiner curvature indeed diverges exactly at the points where the heat capacity diverges. Furthermore, as is shown in Fig.\ref{RRh2D05diag}, the scalar curvature $\mathbb{R}$ always remains nonpositive and negatively diverges, which is analogous to a Van der Waals liquid-gas system. It means that, just as ideal Fermi gases \cite{Oshima:1999rs}, the repulsive interaction is permitted in the microscopic system of the charged AdS black hole incorporating a global monopole.

It is noteworthy that, comparing Figs.\ref{RRh2D05diag} and \ref{RRh2D00diag}, an effect of a global monopole on the microstate of black hole can be deciphered. We find that, due to the presence of global monopole, the divergence process is amazingly shortened, which behaves like the heat capacity in Figs.\ref{CQRhQ3Ddiag} and \ref{CQRh2Ddiag}. A larger $\eta$ corresponds to a shorter unstable process.

%\newpage
\section{Discussions and Conclusions} \label{section6}

Roles of topological defects in spacetime have attracted increasing interest in exploring but are still obscure. This paper focused on further studying the effects of global monopole on charged AdS black hole, involving critical behaviors and microstructure. First of all, for the charged AdS black hole, appearance of a global monopole leads to a strong dependence of thermodynamical quantities on the symmetry breaking parameter $\eta$. Analyzing the stability verified that the charged AdS black hole undergoes a small-large black hole phase transition, of which the critical behaviors qualitatively behaves like a Van der Waals liquid-gas system. Remarkably, differing from the case of the standard RN AdS black hole, we discovered that the critical behaviors of the charged AdS black hole are closely related to the internal global monopole. On the one hand, the critical point of the charged AdS black hole incorporating a global monopole is obviously lowered. On the other hand, parameter $\eta$ have a significant effect on promoting the black hole to reach the stable phase. What's more, microscopic properties of the charged AdS black hole with a global monopole were also discussed by employing the thermodynamic geometry. The result shows that the Ruppeiner scalar curvature shares the same divergent points with the heat capacity. Moreover, the fact that the scalar curvature $\mathbb{R}$ always remains nonpositive and negatively diverges suggests that the charged AdS black hole with a global monopole possesses the microscopic property as ideal Fermi gases. The repulsive interaction is permitted in the microscopic system of the charged AdS black hole. More importantly, in the presence of global monopole, the divergence process of the curvature $\mathbb{R}$ is amazingly shortened. Just as the heat capacity reported, a larger $\eta$ corresponds to a shorter unstable process. As a highlight, comparative illustrations in this work not only contribute to shedding light on the effects of global monopole on the critical behaviors and microscopic properties, but also provide us with valuable clues to the underlying structure of puzzling quantum gravity.

\begin{acknowledgments}\vskip -4mm
The authors would like to appreciate the editor for responsible work at this paper. We also express sincere gratitude to the anonymous referee(s) for high efficiency. This paper is supported by the National Natural Science Foundation of China under Grant Nos.11947086 and 11947046, the Fundamental Scientific Research Foundation of China West Normal University under Grant No.18Q065 and the Doctor Start-up Foundation of Guangxi University of Science and Technology under Grant No.19Z21.
\end{acknowledgments}


\begin{thebibliography}{99}

\bibitem{Vilenkin:1984ib}
A.~Vilenkin,
% Cosmic Strings and Domain Walls,
\textit{Phys. Rept.} {\bf 121}, 263 (1985).
%doi:10.1016/0370-1573(85)90033-X

\bibitem{Vilenkin:1994cs}
A. Vilenkin, E. P. S Shellard, \textit{Cosmic strings and other topological defects} (Cambridge University Press, New York, 2000).

\bibitem{Barriola:1989hx}
M. Barriola and A. Vilenkin,
% Gravitational Field of a Global Monopole,
\textit{Phys. Rev. Lett.} {\bf 63}, 341 (1989).
%doi:10.1103/PhysRevLett.63.341
\bibitem{Deng:2014bta}
G.~M.~Deng,
% Hawking radiation of charged rotating AdS black holes in conformal gravity for charged massive particles, complex scalar and Dirac particles,
\textit{Gen. Rel. Grav.} {\bf 46}, 1757 (2014),
arXiv:1705.04922 [hep-th].
%doi:10.1007/s10714-014-1757-4

\bibitem{Deng:2017abh}
G.~M.~Deng and Y.~C.~Huang,
% $Q$-$\Phi$ criticality and microstructure of charged AdS black holes in $f(R)$ gravity,
\textit{Int. J. Mod. Phys. A} {\bf 32}, 1750204 (2017),
arXiv:1705.04923 [gr-qc].
%doi:10.1142/S0217751X17502049

\bibitem{Addazi:2016prb}
A. Addazi and S. Capozziello,
% The fate of Schwarzschild-de Sitter Black Holes in $f(R)$ gravity,
\textit{Mod. Phys. Lett. A} {\bf 31}, 1650054 (2016),
arXiv:1602.00485 [gr-qc].
% doi:10.1142/S0217732316500541

\bibitem{Chen:2019nsr}
Y.~D.~Chen, W.~Yang and X.~X.~Zeng,
% Thermodynamics and weak cosmic censorship conjecture in Reissner-Nordstr\"{o}m anti-de Sitter black holes with scalar field,
\textit{Nucl. Phys. B} {\bf 946}, 114722 (2019),
arXiv:1901.05140 [hep-th].
%doi:10.1016/j.nuclphysb.2019.114722

\bibitem{Wei:2019ctz}
S.~W.~Wei and Y.~X.~Liu,
% Intriguing microstructures of five-dimensional neutral Gauss-Bonnet AdS black hole,
\textit{Phys. Lett. B} {\bf 803}, 135287 (2020),
arXiv:1910.04528 [gr-qc].
%doi:10.1016/j.physletb.2020.135287

\bibitem{Wei:2019uqg}
S.~W.~Wei, Y.~X.~Liu and R.~B.~Mann,
% Repulsive Interactions and Universal Properties of Charged Anti-de Sitter Black Hole Microstructures,
\textit{Phys. Rev. Lett.} {\bf 123}, 071103 (2019),
arXiv:1906.10840 [gr-qc].
%doi:10.1103/PhysRevLett.123.071103

\bibitem{Jusufi:2020odz}
K.~Jusufi, M.~Azreg-A\"{\i}nou, M.~Jamil, S.~W.~Wei, Q.~Wu and A.~Wang,
% Quasinormal modes, quasiperiodic oscillations, and the shadow of rotating regular black holes in nonminimally coupled Einstein-Yang-Mills theory,
\textit{Phys. Rev. D} {\bf 103}, 024013 (2021),
arXiv:2008.08450 [gr-qc].
%doi:10.1103/PhysRevD.103.024013

\bibitem{Li:2020vpo}
H.~L.~Li, X.~X.~Zeng and R.~Lin,
% Holographic phase transition from novel Gauss\textendash{}Bonnet AdS black holes,
\textit{Eur. Phys. J. C} {\bf 80}, 652 (2020),
%doi:10.1140/epjc/s10052-020-8245-7

\bibitem{Zeng:2020dco}
X.~X.~Zeng, H.~Q.~Zhang and H.~Zhang,
% Shadows and photon spheres with spherical accretions in the four-dimensional Gauss-Bonnet black hole,
\textit{Eur. Phys. J. C} {\bf 80}, 872 (2020),
arXiv:2004.12074 [gr-qc].
%doi:10.1140/epjc/s10052-020-08449-y

\bibitem{Deng:2018wrd}
G.~M.~Deng, J.~Fan, X.~Li and Y.~C.~Huang,
% Thermodynamics and phase transition of charged AdS black holes with a global monopole,
\textit{Int. J. Mod. Phys. A} {\bf 33}, 1850022 (2018),
arXiv:1801.08028 [gr-qc].
%doi:10.1142/S0217751X18500227

\bibitem{Yu:1994fy}
H. W. Yu,
% Black hole thermodynamics and global monopoles,
\textit{Nucl. Phys. B} {\bf 430}, 427 (1994).
%doi:10.1016/0550-3213(94)00339-4

\bibitem{Dadhich:1997mh}
N. Dadhich, K. Narayan, U. A. Yajnik,
% Schwarzschild black hole with global monopole charge,
\textit{Pramana} {\bf 50}, 307 (1998),
arXiv:gr-qc/9703034.
%doi:10.1007/BF02845552

\bibitem{Gao:2002hf}
C. J. Gao, Y. G. Shen,
% De Broglie-Bohm quantization of the Reissner-Nordstr\"{o}m black hole with a global monopole in the background of de Sitter space-time,
\textit{Chin. Phys. Lett.} {\bf 19}, 477 (2002).

\bibitem{Chen:2005vq}
S.~Chen and J.~Jing,
% Late-time behavior of a coupled scalar field in background of a Schwarzschild black hole with a global monopole,
\textit{Mod. Phys. Lett. A} {\bf 23}, 359 (2008),
arXiv:gr-qc/0511098 [gr-qc].
%doi:10.1142/S0217732308023888

\bibitem{Ahmed:2016ucs}
A. K. Ahmed, U. Camci, M. Jamil,
% Accretion on Reissner¨CNordstr\"{o}m¨C(anti)-de Sitter black hole with global monopole,
\textit{Class. Quant. Grav.} {\bf 33}, 215012 (2016),
arXiv:1610.01129 [gr-qc].
%doi:10.1088/0264-9381/33/21/215012

\bibitem{Jusufi:2017lsl}
K. Jusufi, M. C. Werner, A. Banerjee, A.\"{o}vg\"{u}n,
% Light Deflection by a Rotating Global Monopole Spacetime,
\textit{Phys. Rev. D} {\bf 95}, 104012 (2017),
arXiv:1702.05600 [gr-qc].
%doi:10.1103/PhysRevD.95.104012

\bibitem{Rizwan:2018mpy}
A.~Rizwan C.L., N.~Kumara A., D.~Vaid and K.~M.~Ajith,
% Joule-Thomson expansion in AdS black hole with a global monopole,
\textit{Int. J. Mod. Phys. A} {\bf 33}, 1850210 (2019),
arXiv:1805.11053 [gr-qc].
%doi:10.1142/S0217751X1850210X

\bibitem{Das:2019khz}
A.~Das and N.~Banerjee,
% Unitary black hole radiation: Schwarzschild-global monopole background,
\textit{Eur. Phys. J. C} {\bf 79}, 704 (2019),
arXiv:1908.09616 [gr-qc].
%doi:10.1140/epjc/s10052-019-7224-3

\bibitem{Secuk:2019njc}
M.~H.~Se\c{c}uk and \"O.~Delice,
% Superradiance of a global monopole in Reissner-Nordstr\"{o}m(-AdS) space-time,
\textit{Eur. Phys. J. C} {\bf 80}, 396 (2020),
arXiv:1908.00504 [gr-qc].
%doi:10.1140/epjc/s10052-020-7988-5

\bibitem{Ma:2016aat}
Y. B. Ma, R. Zhao and S. Cao,
% Q¨C $\Phi$ criticality in the extended phase space of $(n+1)$ -dimensional RN-AdS black holes,
\textit{Eur. Phys. J. C} {\bf 76}, 669 (2016),
arXiv:1607.00793 [hep-th].
%doi:10.1140/epjc/s10052-016-4532-8

\bibitem{Ruppeiner:1983zz}
G. Ruppeiner,
% Thermodynamic Critical Fluctuation Theory?,
\textit{Phys. Rev. Lett.} {\bf 50}, 287 (1983);

\bibitem{Ruppeiner:1995zz}
G. Ruppeiner,
% Riemannian geometry in thermodynamic fluctuation theory,
\textit{Rev. Mod. Phys.} {\bf 67}, 605 (1995).

\bibitem{Oshima:1999rs}
H. Oshima, T. Obata and H. Hara,
% Riemann scalar curvature of ideal quantum gases obeying Gentile¡¯s statistics,
\textit{J. Phys. A: Math. Gen.} {\bf 32}, 6373 (1999).

\bibitem{Ruppeiner:2018pgn}
G. Ruppeiner,
% Thermodynamic Black Holes,
\textit{Entropy} {\bf 20}, 460 (2016),
arXiv:1803.08990 [gr-qc].
% doi:10.3390/e20060460


\bibitem{Mo:2013sxa}
J.~X.~Mo, X.~X.~Zeng, G.~Q.~Li, X.~Jiang and W.~B.~Liu,
% A unified phase transition picture of the charged topological black hole in Horava-Lifshitz gravity,
\textit{JHEP} {\bf 10}, 056 (2013),
arXiv:1404.2497 [gr-qc].
%doi:10.1007/JHEP10(2013)056

\bibitem{Mo:2015bta}
J.~X.~Mo,
% Weinhold geometry and Ruppeiner geometry of black holes with conformal anomaly,
\textit{Mod. Phys. Lett. A} {\bf 30}, 1550061 (2015).
%doi:10.1142/S0217732315500613

\bibitem{Quevedo:2015fga}
H.~Quevedo, M.~N.~Quevedo and A.~S\'anchez,
% Thermodynamics and geometrothermodynamics of Born-Infeld black holes with cosmological constant,
\textit{Int. J. Mod. Phys. D} {\bf 24}, 1550092 (2015).
%doi:10.1142/S0218271815500923

\bibitem{Soroushfar:2016nbu}
S.~Soroushfar, R.~Saffari and N.~Kamvar,
% Thermodynamic geometry of black holes in f(R) gravity,
\textit{Eur. Phys. J. C} {\bf 76}, 476 (2016),
arXiv:1605.00767 [gr-qc].
%doi:10.1140/epjc/s10052-016-4311-6

\bibitem{L.:2018jss}
C.~L.~A.~Rizwan, A.~Naveena Kumara, K.~V.~Rajani, D.~Vaid and K.~M.~Ajith,
% Effect of Dark Energy in Geometrothermodynamics and Phase Transitions of Regular Bardeen AdS Black Hole,
\textit{Gen. Rel. Grav.} {\bf 51}, 161 (2019),
arXiv:1811.10838 [gr-qc].
%doi:10.1007/s10714-019-2649-4

\bibitem{Soroushfar:2019ihn}
S.~Soroushfar, R.~Saffari and S.~Upadhyay,
% Thermodynamic geometry of a black hole surrounded by perfect fluid in Rastall theory,
\textit{Gen. Rel. Grav.} {\bf 51}, 130 (2019),
arXiv:1908.02133 [gr-qc].
%doi:10.1007/s10714-019-2614-2

\bibitem{Ghosh:2019pwy}
A.~Ghosh and C.~Bhamidipati,
% Thermodynamic geometry for charged Gauss-Bonnet black holes in AdS spacetimes,
\textit{Phys. Rev. D} {\bf 101}, 046005 (2020),
arXiv:1911.06280 [gr-qc].
%doi:10.1103/PhysRevD.101.046005

\bibitem{Yerra:2020oph}
P.~K.~Yerra and C.~Bhamidipati,
% Ruppeiner Geometry, Phase Transitions and Microstructures of Black Holes in Massive Gravity,
\textit{Int. J. Mod. Phys. A} {\bf 35}, 2050120 (2020),
arXiv:2006.07775 [hep-th].
%doi:10.1142/S0217751X20501201

\bibitem{Kumara:2020ucr}
A.~Naveena Kumara, C.~L.~A.~Rizwan, K.~Hegde and K.~M.~Ajith,
% Repulsive Interactions in the Microstructure of Regular Hayward Black Hole in Anti-de Sitter Spacetime,
\textit{Phys. Lett. B} {\bf 807}, 135556 (2020),
arXiv:2003.10175 [gr-qc].
%doi:10.1016/j.physletb.2020.135556

\bibitem{Weinhold:1975mg}
F. Weinhold,
% Metric geometry of equilibrium thermodynamics,
\textit{J. Chem. Phys.} {\bf 63}, 2479 (1975).

\bibitem{Mrugala:1983ct}
R. Mrugala,
% On equivalence of two metrics in classical thermodynamics,
\textit{Physica} {\bf 125A}, 631 (1984).

\end{thebibliography}
\end{document}